\newcommand{\be}{\begin{equation}} \newcommand{\ee}{\end{equation}}
\newcommand{\bea}{\begin{eqnarray}}\newcommand{\eea}{\end{eqnarray}}

\newcommand\s{\scriptscriptstyle}

\documentstyle[12pt]{article}
\begin{document}
\renewcommand{\thefootnote}{\fnsymbol{footnote}}
\begin{titlepage}
\vspace*{5cm}
\begin{center}
 {\Large Harmonic-Superspace Integrability
of  $N=1,  D = 6$ Supersymmetric Gauge Theory}

\vspace{0.5cm}
{\large B.M.Zupnik} \footnote{  E-mail: zupnik@thsun1.jinr.dubna.su}\\
{\it Bogoliubov   Laboratory of Theoretical Physics ,\\
 Joint Institute for Nuclear Research, \\
 Dubna, Moscow Region, 141980, Russia}

\end{center}
\vspace{2cm}
\begin{abstract}
We consider the  harmonic-superspace ($HS$) system of equations that
contains superfield $SYM^1_6$ constraints and  equations of motion. A
dynamical equation in the special $A$-frame is equivalent to the
zero-curvature equation corresponding to a covariant conservation of
$HS$-analyticity. Properties of a general $SYM^1_6$ solution for the
  gauge group $SU(2)$ are studied in the simplest harmonic  gauge.
An analogous approach to the integrability interpretation of
 $SYM$-$SG$-matter systems in $HS$ is discussed briefly.
  \end{abstract}
\end{titlepage}
\renewcommand{\thefootnote}{\arabic{footnote}}
\setcounter{footnote}0
\vspace{1cm}

$\;\;\;$  Integrability interpretations of the supersymmetric $N=3, D=4$
gauge theory $SYM^3_4$ have been considered in the framework of
 twistor-harmonic superspace methods \cite{W1,GIK1}. The on-shell
 superfield constraints of $SYM^3_4$ can be reduced to  integrability
conditions for harmonized (or light-like) spinor covariant derivatives
in the harmonic (twistor) superspace $(HS)$.
Generalized analyticity relations between covariant spinor and harmonic
derivatives arise also in the harmonic formalism \cite{GIK1}. We shall
call a classical field theory  {\it  the harmonic-superspace integrable}
theory if the corresponding dynamic equations are equivalent to the
zero-curvature equations in $HS$. The  harmonic variables play
a role of auxiliary (spectral) parameters for these theories.

The structure of harmonic superspaces for the $HS$-integrable theories
$SYM^3_4$, $SYM^1_{10}$ \cite{W2,Z1} and $SYM^4_5$ \cite{So} is very
complicated.  The simplest harmonic $SU(2)/U(1)$ formalism allows to
 construct the explicit solutions for the self-dual Yang-Mills \cite{GI2}
 and super-Yang-Mills \cite{DeO} theories and also for the non-self-dual
3-dimensional $SYM^6_3$-theory \cite{Z2}.

We propose the $HS$-integrability interpretation of the supersymmetric
$N=1, D=6$ gauge theory $SYM^1_6$ connected via a dimensional reduction
with the $SYM^2_4$-theory. The $HS$-formalism of $SYM^1_6$ has been
 considered in Ref\cite{Z1,HSW} by analogy with the original works on $HS$
\cite{GIK3,GI4}. Our treatment results in a reformulation of the $SYM^1_6$
 superfield constraints and equations of motion as an analytic
 zero-curvature equation plus a linear solvable constraint in $HS$. The
 $HS$-approach produces also an infinite number of conservation laws in
 $SYM^1_6$. We hope that the analogous $HS$-integrability  of $SYM^2_4$
 can help to understand the remarkable quantum properties of this theory
 \cite{SW}.

The off-shell  superfield constraints  for $SYM^1_6$-covariant
derivatives $\nabla^i_a= D^i_a + A^i_a (z)$ in the central basis $(CB)$
 have the following form \cite{Ko,HST}
\be
\{ \nabla^i_a , \nabla^k_b \} + \{ \nabla^k_a , \nabla^i_b \}= 0
\label{A1}
\ee
where $D^i_a$ is a plane spinor derivative and $A^i_a (z)$ is a
$CB$-gauge spinor superfield.
(We consider  the $SO(5,1)$-spinor indices $a, b \dots=1\ldots 4 $
and the $SU(2)$-spinor indices $i,k\ldots=1,2$ and use $N=1, D=6$
superspace coordinates $z=x^{ab},\theta^a_i$.)
The  harmonic formalism with $SU(2)/U(1)$ harmonics $u^{\pm}_i$ leads to
the $HS$-integrability interpretation of these constraints
\be
\{ \nabla^+_a , \nabla^+_b \} = 0\;\Rightarrow\;\nabla^+_a =u^+_i\;
\nabla^i_a = u^+_i\;h^{-1}\;D^i_a h
\label{A2}
\ee
where $h$ is an off-shell 'bridge' matrix \cite{GIK3}.

The $SYM^1_6$ superfield equation of motion in $CB$ has a dimension
$d=-2$ in the  units of length
\be
\nabla^i_a W^{ak} + \nabla^k_a W^{ai}= 0,\hspace{1.5cm} W^{ak}=(i/12)
 \varepsilon^{abcd}\varepsilon_{jl}[\nabla^k_b\;,\{\nabla^j_c\;,
\nabla^l_d\}]
 \label{A3}
\ee
where $W^{ai}  $ is
a covariant superfield containing the field-strength multiplet of
$SYM^1_6$. This independent nonlinear equation is the 2-d part of the
 $SYM^1_6$-system of equations.

            A harmonic (twistor) transform  from $CB$ to the
analytic basis $(AB)$ of $SYM^1_6$ can be defined in terms of the bridge
matrix $h$ . Consider the basic $d=0$ harmonic equation of the $V$-frame
 in $AB$ \cite{GIK3}
\be
 (\partial^{\s++} + V^{\s++}) h(z,u)=\nabla^{\s++} h =0
 \label{A4}
\ee
 where a harmonic derivative
$\partial^{\s++}$  is used.  The off-shell prepotential
$V^{\s++}(\zeta,u)$ lives in the analytic $N=1, D=6$ superspace $AHS^1_6$.
 The natural coordinates of $AHS^1_6$ are $\zeta= (x_{\s A}^{ab},\;
\theta^c_+ )$ \cite{Z1,HSW} but one can use the central coordinates $z,u$,
 too.

The   $HS$ action, basic constraints and equation of motion for $SYM^1_6$
have been written in Ref\cite{Z1}.
The basic $V$-frame off-shell constraints are:

1) The harmonic-zero-curvature $(HZC)$ equation (d=0) that contains the
harmonic connections $V^{\s++} , A^{\s--}$
\be
\partial^{\s++}_{\s A} A^{\s--}
 - \partial^{\s--}_{\s A} V^{\s++} + [V^{\s++} , A^{\s--}] = 0
\label{A5}
\ee
and can be treated as an equation for $A^{\s--}(V^{\s++})$.
 One can use the  perturbative or nonperturbative solution of
 this integrable equation \cite{Z1,Z2}.

 2)The 'kinematic' $V$-analyticity constraint \cite{GIK3}
 \be
 [\nabla^+_a,\nabla^{\s++}]= D^+_a\;V^{\s++}=0,\hspace{1.5cm}
D^+_a=u^+_i\;D^i_a
 \label{A5a}
 \ee

 3) The  conventional constraint for the spinor
 connection $A^-_a$ \cite{Z1}
 \be
 [\nabla^{\s--},\;\nabla^+_a  ] =\nabla^-_a =D^-_a + A^-_a=
 D^-_a - D^+_a A^{\s--}
 \label{A5b}
  \ee

 4) The initial $CB$ constraint (\ref{A2}) transforms to the trivial
 $AB$ constraint \cite{GIK3}
 \be
  \{\nabla^+_a,\nabla^+_b \} = \{D^+_a,D^+_b \} = 0
 \label{A5c}
 \ee

Consider the $SYM^1_6$ equation of motion in the $V$-frame
\be
F^{\s++}=(1/4)D^+_a\;W^{a+}(V)=(1/24)\varepsilon^{abcd}\;D^+_a\;
D^+_b\;D^+_c\;D^+_d\;A^{\s--}(V) = 0
\label{A6}
\ee
  Stress that $A^{\s--}(V)$ is a nonlinear
harmonic functional of $V^{\s++}$ so it is difficult to analyze
this equation directly in the $V$-frame.

Let us introduce a new  $A$-frame which uses $A^{\s--}(z,u)$ as a basic
on-shell superfield. The whole $A$-system of $SYM^1_6$-equations  for
covariant derivatives with $d\ge -2$ is identical to the corresponding
 $V$-system, however we change {\it the interpretation, the basic set
of equation and the auxiliary-field structure } of the harmonic formalism
in the new frame.
 A basic $A$-bridge equation   contains the covariant derivative
 $\nabla^{\s--}=\partial^{\s--} + A^{\s--}$ (see below),
 and the bridge equation (\ref{A4}) become a secondary equation.
We preserve the representation (\ref{A2}) for the spinor $CB$-connection,
the conventional constraint (\ref{A5b}) and the basic relation of $AB$
(\ref{A5c}).
The $HZC$-equation (\ref{A5}) in the $A$-frame is treated as an
 integrable equation for the connection $V^{\s++}(A^{\s--})$.
 Harmonic equations of a dimension $d=0$
  do not guarantee the conservation of analyticity. The kinematic
analyticity equation (\ref{A5a}) of the $V$-frame transforms to a
dynamical equation of the $A$-frame for the solution $V^{\s++}(A^{\s--})$
 of the $HZC$-equation. One can use also the equivalent dynamical
 analyticity condition for the covariant derivatives $\nabla^-_a$ and
 $\nabla^{\s--}$. The analyticity equations play a role of the
 $HS$-integrability conditions of the $SYM^1_6$-theory.

 The nonlinear in $V^{\s++}$ equation (\ref{A6}) corresponds to a simple
 linear $d=-2$ constraint of the $A$-frame. This constraint has the
 following general solution:
\be
 A^{\s--}(z,u) = D^+_a\;A^{a\s(-3)}(z,u)
\label{A7}
\ee
where $A^{a\s(-3)}$ is the on-shell $SYM^1_6$ prepotential . The gauge
transformation of this prepotential contains a term producing the standard
$AB$-gauge transformation $\delta A^{\s--} = \nabla^{\s--}\lambda$
and an additional term with the spinor derivative of a general symmetric
 spinor.

The bridge matrix $h_{\s A}=h(A^{\s--})$ of the $A$-frame is a solution of
  the following harmonic equation
\be
\nabla^{\s--}\;h_{\s A} = (\partial^{\s--}_{\s A} + D^+_a\;A^{a\s(-3)}\;)
\;h_{\s A} = 0
\label{A16}
\ee
This equation on the sphere $SU(2)/U(1)$ is a harmonic part of the
linear problem for the $HS$-integrable $SYM^1_6$-system of equations.
 A key point of the harmonic approach is an integrability of the harmonic
bridge and $HZC$ equations. We shall consider
 below an explicit solution for the $SU(2)$ gauge group .

 Thus, a transition to the new frame simplifies significantly the
 $SYM^1_6$-system of equations and allows to solve the equation with
 $d=-2$. Now we can introduce a dynamical zero-curvature equation
  $( AZC$-equation ) of the $A$-frame
\be
[ \nabla^-_a ,\;\nabla^{\s--} ] =D^-_a\;D^+_b\;A^{b\s(-3)} +
\partial^{\s--}\;D^+_a\;D^+_b\;A^{b\s(-3)}
- [D^+_a\;D^+_b\;A^{b\s(-3)},\;D^+_c\;A^{c\s(-3)}]=0 \label{A9}
\ee
where  the conventional constraint (\ref{A5b}) and the representation
 (\ref{A7}) are used. This equation is treated as the basic dynamical
equation of the $SYM^1_6$-theory.

The $AZC$ integrability condition is equivalent to  component $SYM^1_6$
equations of motion. This can be proved in
a normal gauge for the prepotential $A^{a\s(-3)}$
that contains an infinite number of harmonic
 auxiliary fields  and the physical component $SYM^1_6$ fields:
 the vector $A_{ab}(x_{\s A})$, the spinor $\psi^a_i(x_{\s A}) $
 and the independent field-strength $F^a_b(x_{\s A})$ . All
 auxiliary fields vanish in a consequence of the $AZC$-equation, and the
 standard first-order component $SYM^1_6$ equations for $F, A$ and $\psi$
 arise, too. It is evident that the $A$-frame and the $V$-frame are
equivalent on-shell and have identical component solutions for the
physical fields.

An alternative $VZC$-formulation of the $HS$-integrability condition in
the $A$-frame has the following form:
\be
[\nabla^+_a ,\;\nabla^{\s++} ]= D^+_a V^{\s++}(A^{\s--}) =0 \label{A12}
\ee
where the $V$-solution of $HZC$-equation (\ref{A5}) and the representation
(\ref{A7}) should be used.
This form of the dynamical $SYM^1_6$-equation is equivalent to
Eq(\ref{A9}). Below we shall discuss   the $HS$-integrability
 condition for the gauge group $SU(2)$.

 All known integrable field theories have an infinite number of
 conservation laws. The explicit construction of the conserved quantities
 follows immediately from the zero-curvature representation and has a
clear geometric interpretation in terms of the contour variables . For
 example, the explicit constructions of the conservation laws  in the
 twistor formalism of $SYM^3_4$ have been discussed in Ref\cite{De}.

Let us choose a time variable $t=x^{12}$ and introduce the corresponding
covariant derivative
\be
\nabla_t = \nabla_{12} = \partial_t + A_{12}= \partial_t +i D^+_1\;
D^+_2\;A^{\s--} \label{A15}
\ee
The basic spinor $AZC$-equation (\ref{A9}) generates the important
 relation $[\nabla^{\s--},\;\nabla_t] = 0 $.

It should be stressed that the bridge $h_{\s A}$ is a natural harmonic
 analogue of the  contour variables in the zero-curvature representation.
 A covariant constancy of the bridge in  all spinor and vector directions
 is a consequence of the dynamic analyticity equation (\ref{A9}), for
 instance, we have the on-shell condition of a covariant
 conservation $\nabla_t\;h_{\s A} = 0$ where the $t$-component of the
$CB$-connection should be used, too.  The simplest conserved quantities
  can be constructed as invariant functions of $h_{\s A}$ by analogy with
 the integrable $\sigma$-models.

The B\"acklund  transformations ($BT$) play an important role for
integrable theories as transformations in the spaces of solutions.
For the self-dual $SYM$-theories  these transformations have been
 considered in the special gauge \cite{DeL}. We shall discuss $BT$ in
 the $HS$-formalism of $SYM^1_6$.

Let $A^{\s--}$ and $\hat{A}^{\s--}$ be two different solutions of
$SYM^1_6$-system (\ref{A9}). Consider the corresponding bridges $h_{\s A}$
and $\hat{h}_{\s\hat{A}}$. Then the B\"acklund transformation between
 these solutions has the  form of a nonanalitic gauge transformation with
 a gauge matrix $B(A,\hat{A})= h_{\s A}\; \hat{h}_{\s\hat{A}}^{-1}$.

 The $HS$-integrability interpretation allows us to analyze the
explicit constructions of the $SYM^1_6$-solutions by analogy with the
harmonic formalism of $SDSYM$ \cite{DeO} or $SYM^6_3$ \cite{Z2} theories.
Let us go back to the $V$-frame and consider $SYM^1_6$-system
in the case of the gauge group $SU(2)$. We shall use a harmonic
 representation of the general $SU(2)$ prepotential \cite{Z2}
 \footnote{An analogous representation was used also in Refs
 \cite{KS,O} for the harmonic analysis of $SDYM$-theory.}
\be
V^{\s++} = (U^{\s+2})\; b^0 (\zeta,u) + (U^0)\;b^{\s(+2)} (\zeta,u) +
  (U^{\s-2})\; b^{\s(+4)}(\zeta,u) \label{A21}
\ee
where $b^{0},\;b^{\s(+2)},\;b^{\s(+4)}$ are arbitrary real analytic
superfields
and $(U^{q})$ are matrix generators of the Lie algebra $SU(2)$ in a
 harmonic representation
\be
 (U^{\s\pm 2})^i_k = u^{\pm}_k \; u^{\pm i} ,\;\;\;\; (U^0)^i_k =
 u^{-}_k\;u^{+i} + u^{+}_k\;u^{-i}  \label{A21a}
 \ee

We can obtain the simplest general gauge for the $SU(2)$-prepotential
using gauge transformations of the harmonic components $b^{(q)}$\cite{Z2}
\be
V^{\s++}(b^0,\rho) = (U^{\s+2})\; b^0 (\zeta,u) + (U^{\s-2})\;
(\theta_+)^4\;\rho  \label{A22}
\ee
where $b^0$ is an arbitrary analytic function and $\rho$ is some constant
(a 'vacuum' field) which characterizes the choice of different phases
 in $SYM^1_6$-theory. Remark that the corresponding harmonic-independent
 term in $b^{\s(+4)}$ contains a trace component of the auxiliary matrix
scalar field with $d=-2$ and can be written as
 $D^{ik}_{ik}(x)=\rho + \partial^{ab} f_{ab}(x_{\s A})$.

 This $(b^0,\rho)$-gauge has the non-Abelian residual $\lambda$-gauge
 invariance with an arbitrary component $\lambda^{\s(-2)}$ and restricted
 components $\lambda^{\s(+2)},\;\lambda^0$ \cite{Z2}.

The harmonic bridge equation (\ref{A4}) with the prepotential
$V^{\s++}(b^0,\rho) $ can be integrated in quadratures. The integration
procedure uses a nilpotency of the last term in Eq(\ref{A22} ). We do not
discuss here the explicit solution for $h(b^0,\rho)$ and restrict
ourselves to the partial choice $\rho=0$.

The phase of $SYM^1_6$ and $SYM^2_4$ with $\rho=0$ has been considered in
Refs\cite{Z2,ZT}, and the harmonic $V$-frame equations have been solved
 in quadratures.

Now we shall discuss properties of the $SU(2)$-solution in the alternative
$A$-frame. The first step of this approach is a solution of  harmonic
equations in the representation (\ref{A7}) and then the dynamic
analyticity equation should be solved.

Consider the $A$-frame solution of the bridge equation (\ref{A16})
\be
h_{\s A}= \mbox{exp}[(1/2)(U^0)\;\mbox{ln}\;(1-a^0 )][1 - (U^{\s+2})\;
a^{\s(-2)}]
\label{A30}
\ee
\be
a^0 (z)=D_{(a}^+D_{b)}^-\;A^{ab}(z),\;\;\;\;a^{\s(-2)}(z,u)=D^+_a\;
A^{a\s(-3)}(z,u)
\label{A31}
\ee
where parentheses denote a symmetrization of indices. The $(U^{q})$
 components of the connection can be obtained from the formula
 $A^{\s--}=h_{\s A}\partial^{\s--}h_{\s A}^{-1}$. Note that this solution
has the singular point\footnote{This singularity corresponds to
the irregular $V$-frame prepotential with $b^0=-1$ \cite{Iv}} $a^0=1$.

This formula looks as some Ansatz in the $A$-frame which partially uses
information from the $V$-frame. Nevertheless, this way produces an
alternative method to analyze the general $SU(2)$-solution in the
$(b^0,0)$-gauge since all frames are equivalent on-shell.

The corresponding harmonic connection $V^{\s++}(A)$ has the following
 $A$-frame components:
\be
b^0 (A)=\frac{a^0 + \partial^{\s++}a^{\s(-2)}}{1-a^0},\;\;\;
b^{\s(+2)}(A)=0,\;\;\;b^{\s(+4)}(A)=0 \label{A32}
\ee
The analyticity condition $D^+_a\;b^0 (A)=0$ produces the dynamical
equation of this approach. One can write this equation in a polynomial
 form
\be
(1 + \partial^{\s++}a^{\s(-2)})\;D^+_a\;a^0 + (1-a^0 )\;D^+_a\;
\partial^{\s++} a^{\s(-2)} = 0 \label{A33}
\ee

Note that this one-component equation is covariant under the non-Abelian
gauge transformations of the harmonic gauge.
Thus, the $SYM^1_6$-system of equations reduces to Eq(\ref{A33}) in the
  $(b^0,0)$-gauge. This reduction simplifies significantly
 the initial $SYM^1_6$-system and gives the hope to obtain the explicit
solutions of this problem.

The residual gauge invariance of the $(b^0,0)$-gauge can be fixed in
the $V$-frame, and the function $b^0$ (\ref{A22}) can be written in terms
 of some $d=2$ prepotential \cite{ZT}. The classical action of $SYM^1_6$
 in this gauge has the following form:
\be
S(b)=\frac{1}{g^2}\int d^{14}z [\mbox{ln}(1+b(z))-b(z)] \label{A37}
\ee
\be
b(z) = \int b^0 (z,u) = D^{iklm}_{(4)} \;V_{iklm}(z) \label{A28}
\ee
where  $D^{iklm}_{(4)}$ is the 4-th order combination of $D^i_a$
\cite{Ko}.

A spinor component of the gauge $CB$-superfield   can be written also
  in terms
of the single superfield $b(z)$ \cite{ZT}
\be
\left[A^l_a(z)\right]^k_i = \frac{1}{1+b(z)}\left[\delta^l_i D^k_a b(z) -
(1/2)\delta^i_k D^l_a b(z)\right] \label{A34}
\ee

The harmonic-superspace integrability of the $SYM^1_6$-theory
guarantees an analogous property of its $4D$ subsystem $SYM^2_4$.
Consider the representation (\ref{A7}) in the Euclidean version of
$SYM^2_4$
\be
A^{\s--}(z,u)=D^+_\alpha \;A^{\alpha\s (-3)} + \bar{D}^+_{\dot{\alpha}}\;
\bar{A}^{\dot{\alpha}\s(-3)} \label{A35}
\ee
where the two-component $SU_L(2)$ and $SU_R(2)$ spinors are used.

The case $A^{\alpha\s (-3)}=0$ corresponds to the general self-dual
 solution of $SYM^2_4$
\be
W(A)=(\bar{D}^+)^2\;A^{\s--} = 0 \label{A36}
\ee
The self-dual prepotential $\bar{A}^{\dot{\alpha}\s (-3)}$ satisfies also
the nonlinear $AZC$-equation (\ref{A9}). Specific features of
 $SYM^2_4$-solutions will be discussed elsewhere.

 The effective quantum action of $SYM^2_4$\cite{SW}
 can be rewritten in terms of $N=2$ superfields. It seems natural to
suppose that the classical $HS$-integrability of $SYM^2_4$ is connected
with the remarkable quantum structure of this theory. An analysis
of the quantum properties of the nonrenormalizable $SYM^1_6$-theory
is a more difficult problem. Note that the simplest harmonic gauge
 condition for the gauge group $SU(3)$ conserves the analytic
components $b^0_3$ and $b^{(+2)}_8$ corresponding to the Cartan
generators of $SU(3)$ and the constant $\rho$-term (\ref{A22}). The
 analogous harmonic gauges can be founded for any gauge group.

The  integrable theory $SYM^3_4$ can be described in the framework of
 $SYM^2_4$ with the special hypermultiplet interactions \cite{GI4}. An
analogous construction exists for  the $SYM^2_6$-theory in
terms of $HS^1_6$-superfields. It seems useful to consider the
$A$-frame $HS$-equations of  more general interacting
$SYM$-supergravity-matter systems. Any $HS$-integrable system can be
 reduced to the dynamical analyticity conditions and some solvable linear
 constraints. This formulation may help to build the
explicit classical solutions and to study quantum solutions.

In conclusion one should stressed that the transition to the new
$A$-frame of the $HS$-formalism allows us to prove the integrability
of the $SYM^1_6$- and $SYM^2_4$-theories. We hope that a manifest
 supersymetry of the harmonic formalism, the $HS$-integrability and the
use of the special harmonic gauges can simplify significantly the study
of these theories.

The author would like to thank cordially E.A.Ivanov and V.I.Ogievetsky
for stimulating discussions and critical remarks and  C.Devchand,
A.A.Kapustnikov, A.D.Popov, K.S.Stelle and M.A.Vasiliev for discussions.
I am grateful to A.T.Filippov and the administration of LTP JINR for
support and hospitality. This work is partially supported by ISF-grant
 RUA000 and INTAS-grant 93-127.

\end{document}